\documentclass[12pt,letterpaper]{article}
\usepackage{amsmath,amssymb,pgf,pgfarrows,pgfnodes,float,appendix, hyperref}
\usepackage{graphicx}
\usepackage{subfigure}
\usepackage[margin=0.9in]{geometry}

\newcommand{\be}{\begin{equation}}
\newcommand{\ee}{\end{equation}}
\newcommand{\bea}{\begin{eqnarray}}
\newcommand{\eea}{\end{eqnarray}}

\title{{\rm\footnotesize \qquad \qquad \qquad \qquad \qquad \ \qquad \qquad \qquad \ \ \ \ \ \                  RUNHETC-2016-15, UTTG-12-16 }\vskip.5in    Schrodinger's Cat and World History:The Many Worlds Interpretation of Alternative Facts  }
\author{Tom Banks\\
Department of Physics and NHETC\\
Rutgers University, Piscataway, NJ 08854\\
E-mail: \href{mailto:banks@physics.rutgers.edu}{banks@physics.rutgers.edu}
\\
\\
}
\date{April 1, 2017}
\begin{document}
\maketitle

\begin{abstract}
 I propose that much recent history can be explained by hypothesizing that sometime during the last quarter of 2016, the history of the world underwent a macroscopic quantum tunneling event, creating, according to the Many Worlds Interpretation, a new branch of the multiverse in which my consciousness and that of my readers is now trapped.  The failure of much political polling is then understood by assuming that the particular branch we are on had very low amplitude in the quantum wave function of the multiverse. In this view, one must take a different attitude towards alternative facts than that proposed by the mainstream media. We know that quantum tunneling can change the low energy laws of physics in the different branches of the wave function.  Alternative facts may simply be the reflection of the media's ignorance of the state of the world after a quantum transition of this magnitude.
\end{abstract}

\section{Introduction}

According to the Copenhagen interpretation of Quantum Mechanics (QM) the quantum wave function is simply a device for computing probabilities. Advocates of the Many Worlds Interpretation instead want to view the quantum state as a physical object, evolving deterministically.  

Consider the famous Schrodinger's Cat experiment.  A cat is put in a gas chamber and the lethal gas valve is either opened or left closed, depending on the polarization of the spin of some elementary particle.  When the particle's spin is in a superposition of UP and DOWN states, the entanglement of this state with the macroscopic state of the gas chamber/cat puts the cat into a superposition of being alive and dead.  The conventional Copenhagen interpretation of this is that the quantum state is just a probability distribution, and that, because of decoherence, the quantum probabilistic prescriptions for macroscopic variables obey Bayes' rules for conditional probabilities with incredible accuracy.  The superposition of live and dead cats is just the statement that the theory does not predict whether the cat will be alive or dead after the entanglement event, but only the probabilities of one or the other of those alternatives.

Adherents of the MWI claim instead that the quantum state of such a macroscopic superposition represents a true branching of the world into two parallel worlds
which no longer communicate\footnote{The tiny interference effects, which violate Bayes' conditional probability rules, are the vestiges of communication between the two worlds.}.  The Schrodinger equation for the time evolution of the quantum state represents a real physical process.  The quantum amplitudes of different terms in a superposition have a status similar to that of the value of a field at different points, and their interpretation in terms of probabilities is supposed to be derived, rather than postulated.

The Copenhagen and MW interpretations of QM are supposed to make identical predictions for experiments, so there seems to be no scientifically sound procedure for differentiating between them.  Adherents of each will give cogent philosophical arguments in favor of their preferred interpretation, and talk past each other, much like partisan politicians. Until recently, there seemed to be no way to distinguish between the two.  This note will argue that recent political history makes a decisive experimental case for the MWI.

\section{Macroscopic Quantum Tunneling}

Suppose you find yourself at the bottom of a deep well, with a limited food supply.  There is literally not enough energy available in your body plus its food supply, to enable you to climb out of the gravitational pull of the earth, which keeps you pinned to the bottom of the well.

The preceding paragraph was a description of your plight according to classical physics.  In QM it is incorrect.  According to QM, the center of mass of your body cannot be localized at the bottom of the well, and have a definite energy as well.  Your quantum state thus predicts that there is a finite probability that your body will be found outside the well.  This sort of  apparent violation of conservation of energy is called {\it quantum tunneling}.    
The probability for this occurring for a macroscopic object is so tiny, that it has been deemed impossible to 
test this prediction of QM by an actual experiment.

Modern quantum field theory seems to predict an even more dramatic type of event, called vacuum tunneling or vacuum decay\cite{cdl}
in which the laws of physics and chemistry undergo a dramatic shift.  To quote one of the seminal papers on the subject 
\begin{itemize}
\item The possibility that we are living in a false vacuum has never been a cheering one to contemplate. Vacuum
decay is the ultimate ecological catastrophe; in a new vacuum there are new constants of nature; after vacuum decay, not only is life as we know it impossible, so is chemistry as we know it. However, one could always draw Stoic comfort from the possibility that perhaps in the course of time the new vacuum would sustain, if not life as we know it, at least some structures capable of knowing joy, This possibility has now been eliminated.
\end{itemize}
These ideas have become quite popular with the advent of string theory, where the vast landscape of possible descriptions of the laws of physics are interpreted as different basins of attraction in a vast multiverse, which are connected by vacuum tunneling transitions.

The description of vacuum tunneling is quite interesting. It begins with a microscopic event, which takes place over a microscopically small time scale in a microscopic region of space.  This ``true vacuum bubble" then begins to grow rapidly, reaching the speed of light in a microscopic time.  Observers outside the bubble are quickly engulfed by it.

There is no real theory of what the different possibilities are for basins of attraction and tunneling events connecting them.  It's generally been assumed that the new laws of physics are so dramatically different that no observer could survive the transition, so there has been no attempt to investigate what the transition would ``feel like".

According to the MWI, the different universes accessed by vacuum tunneling are equally real, and can only communicate with each other by highly improbable tunneling events. Observers in different basins of attraction can never communicate with each other, so there seemed to be no way, even in principle, to test this theory.

The present author has been very skeptical of this whole set of ideas, both the String Landscape and the Many Worlds Interpretation of Tunneling\cite{landskepticism}.  The MWI in particular seemed to be a completely unecessary overlay on the statistical intepretation of QM, completely untestable by experiment.  However, the events that have transpired, beginning sometime in 2016 on the planet Earth, have led me to re-evaluate this viewpoint, in the face of what appears to be experimental confirmation of a version of Landscape/MWI theories.

Let me make it perfectly clear that the observational evidence does not go so far as to completely confirm those theories in all detail.  In fact, that evidence suggests that at least the hypothesis of the standard MWI scenario, according to which the new universe would be so different from our own that we could not survive the bubble collision, is wrong.

Rather, the new universe is similar in many respects to our own, except for a few improbable events, which change the course of history.  Furthermore, the collision with the bubble wall that engulfed us appears to have only psychological rather than dramatic physical effects.

In short, the hypothesis of this paper is that the improbable result of the election for the Presidency of the United States, was the result of a low amplitude tunneling event in the wave function of the universe. This simple hypothesis explains at a stroke why the highly scientific statistical methodologies of researchers like 538.com and the Princeton Election Consortium failed to predict this occurrence.  It also explains how experienced pundits like Joe Scarborough could have miscalled the election.  Both the intuitive models of the pundits and the mathematical models of the pollsters are based on classical models of the political process.  The erratic fluctuations in the behavior of one of the presidential candidates cannot be understood in such a model and require instead a model of that candidate as a superposition of different personalities.  This is a manifestation of the non-classical nature of the quantum state of a macroscopic object DURING a tunneling transition.  According to conventional Copenhagen quantum mechanics, such superpositions are unobservable since the wave function of a candidate represents only statistical predictions of what his behavior MIGHT be, rather than actual events in the world.  Their direct manifestation is {\it prima facie} evidence in favor of the MWI and a refutation of the Copenhagen interpretation of QM.

The highly improbable tunneling event, which caused this transition is unlikely to recur, and so studying it in detail may be our only chance to finally understand the meaning of quantum mechanics.  Since we are unlikely to see a recurrence of a tunneling event in the history of our planet, I have decided to call this one The Last Trump.

We will present evidence that the space-time origin of
The Last Trump was at approximately [XXXXXX|\footnote{The Department of Homeland Security/Ministry of State Security has judged that certain of the passages in this paper might be prejudicial to national security.  Following the guidelines of Executive Order 474,543.06, these passages have been deleted from all copies of this document.  Possession of unexpurgated copies is a Federal felony under Executive Order 575,666.37, punishable by imprisonment in a Black Site for a period of undetermined length.} in the city of M[XXXXX] in the R[XXXX] F[XXXX]. The detailed evolution of the wall of expanding Trump can be followed by a close examination of reliable news sources during the course of 2016. The passage of the bubble wall can be gauged by calculating the ratio of real to fake news about the political situation at each space-time location.  This procedure is however subject to the following caveat: after a region has been engulfed by the expanding bubble, it is in the new vacuum, where the laws of nature MAY have changed.  We can't know what is true in the new branch of the multi-verse, without redoing centuries of experimental work in all of the sciences, including the social, political and journalistic sciences.  Thus, the definition of real and fake news may well be different before and after the passage of the bubble wall.  

Mainstream media have derided the ``alternative facts" presented by members of the current administration to justify their actions.  In the MW Interpretation of recent history, it is equally likely that it is the nature of truth itself which has changed. The mainstream media, in clinging to a now obsolete notion of truth, appropriate for a bygone universe, which is unlikely to recur\footnote{See however the last footnote of the Conclusions.}, are doing a disservice to the public. 
Their real job now is to do research into what the new nature of truth actually is.  They should probably shut down their publishing operations and devote full staff time to researching the nature of Trump's universe, or reserve the published reports for material supplied by the administration.  

In the next section, I will present detailed evidence for all of these claims.  This will be followed by a short set of Concluding Remarks.

\section{The Evidence}

[XXXXXXXXXXXXXXXXXXXXXXXXXXXXXXXXXXXXXXXXXXXXXXXXXXXXXXXXXXXXXXXXXX
XXXXXXXXXXXXXXXXXXXXXXXXXXXXXXXXXXXXXXXXXXXXXXXXXXXXXXXXXXXX
XXXXXXXXXXXXXXXXXXXXXXXXXXXXXXXXXXXXXXXXXXXXXXXXXXXXXXXXXXXXX
XXXXXXXXXXXXXXXXXXXXXXXXXXXXXXXXXXXXXXXXXXXXXXXXXXXXXXXXXXXXXX
XXXXXXXXXXXXXXXXXXXXXXXXXXXXXXXXXXXXXXXXXXXXXXXXXXXXXXXXXXXXX
XXXXXXXXXXXXXXXXXXXXXXXXXXXXXXXXXXXXXXXXXXXXXXXXXXXXXXXXXXX
XXXXXXXXXXXXXXXXXXXXXXXXXXXXXXXXXXXXXXXXXXXXXXXXXXXXXXXXXX]

\section{Conclusions}

Open minded readers, uncorrupted by the Enemies of the People, will have been convinced by the extensive scientific data presented in the previous section, that this article is correct\footnote{Note Added by DHL/MSS: A pending Executive order, tentatively numbered 2,437,456.0098, will make contrary opinions subject to as yet unspecified sanctions (But they'll be bad!!!).}. The Copenhagen interpretation of QM is wrong, MWI is correct, Alternative Facts are Truth, the old notions of Truth and the outmoded Scientific Method were features of the pre-tunneling, UNSTABLE, universe, AND WILL NEVER BE USEFUL AGAIN. President Trump will make America Great and we will dwell in The Last Trump for ever\footnote{I should note that in previous papers\cite{cdlx}, the author argued that many improbable vacuum tunneling transitions, were actually transitions to short lived meta-stable states. These states often end in a catastrophic transition back to the original stable vacuum. The current universe's version of this author denies the validity of those arguments. Readers familiar with the work of Sinclair Lewis will understand the meaning of this footnote.}.

 \vskip.3in
\begin{center}
{\bf Acknowledgments }\\
The work of T.Banks is {\bf\it NOT} supported by the Department of Energy, since that Department, and all other Departments of the Federal Government not directly concerned with Security or Making American Great will soon be eliminated.
\end{center}

\end{document}